# A Hybrid Multi-Objective Particle Swarm Optimization Method to Discover Biclusters in Microarray Data

Mohsen lashkargir *
Department of Computer Engineering
Islamic Azad University, najafabad branch
Isfahan, 81746, Iran
e-mail: mlashkargir@gmail.com

S. Amirhassan Monadjemi
Department of Computer Engineering
Faculty of Engineering
University of Isfahan
Isfahan, 81746, Iran

Ahmad Baraani Dastjerdi
Department of Computer Engineering
Faculty of Engineering
University of Isfahan
Isfahan, 81746, Iran

*Abstract* — In recent years, with the development of microarray technique, discovery of useful knowledge from microarray data has become very important. Biclustering is a very useful data mining technique for discovering genes which have similar behavior. In microarray data, several objectives have to be optimized simultaneously and often these objectives are in conflict with each other. A Multi-Objective model is capable of solving such problems. Our method proposes a Hybrid algorithm which is based on the Multi-Objective Particle Swarm Optimization for discovering biclusters in gene expression data. In our method, we will consider a low level of overlapping amongst the biclusters and try to cover all elements of the gene expression matrix. Experimental results in the bench mark database show a significant improvement in both overlap among biclusters and coverage of elements in the gene expression matrix.

## I. INTRODUCTION

The microarray technique allows measurement of mRNA levels simultaneously for thousands of genes. It is now possible to monitor the expression of thousands of genes in parallel over many experimental conditions (e.g., different patients, tissue types, and growth environments), all within a single experiment. Microarray data constructs a data matrix in which rows represent genes and columns show condition. Each entry in the matrix shows the expression level of specific gene ($g_i$) under particular condition ($c_i$). Through analysis of gene expression data the genes are found that represent similar behavior among a subset of condition. In [14] used clustering for analyses of gene expression data but genes don't show similar behavior in all conditions, while genes show similar behavior in subset of conditions. However the genes are not necessarily related in all conditions, in other words, there are genes that can be relevant in subset of condition [3]. In fact, both of rows and columns (genes and conditions) are clustered and they refer to biclustering (simultaneously clustering of both rows and columns)[2].

The biclustering problem is even more difficult than clustering, as we tried to find clusters using two dimensions, instance of one.

The first biclustering useful algorithm was proposed by Cheng and Church [1] in 2000. They introduced the residue of an element in the bicluster and the *mean squared residue* of submatrix for quality measurement of biclusters. This introduced method is a good measurement tool for biclustering and we use this measurement. Getz et al [15] presented the couple two-way clustering. It uses hierarchical clustering applied separately to each dimension and they define the process to combine both results. The time complexity of this method is Exponential. Yang improved Cheng and Church approach to find K possibly overlapping biclusters simultaneously [3]. It is also robust against missing values which are handled by taking into account the bicluster volume (number of non-missing elements) when computing the score.

The biclustering problem is proven to be NP hard [1]. This high complexity motivated the researcher to use stochastic approach to find biclusters. Federico and Aguilar proposed a Biclustering algorithm with Evolutionary computation [4]. In biclustering of gene expression data, the goal is to find bicluster of maximum size with mean squared residue lower than a given δ, which are relatively high row variance. In [4], the fitness function is made by the sum weighted of this objectives function. Since in biclustering problem some objectives exist, that are in conflict with each other, using multi object methods is very suitable to solve that. In this work we address a biclustering problem with Multi-Objective problem that hybrid with Cheng and Church algorithm.

This paper is organized as follows: in section 2, the definitions related to biclustering are presented. An introduction to PSO and Binary PSO is given in section 3. The

                                                                                                                         

description of the algorithm is illustrated in section 4. Experimental results and comparative analysis are discussed in section 5. The last section is the conclusion.

## II. BICLUSTRING

A bicluster is defined on a gene expression matrix. Let $G=\{g_1, \ldots, g_N\}$ be a set of genes and $C=\{c_1, \ldots, c_M\}$ be a set of conditions. The gene expression matrix is a matrix of real numbers, with possible null values, where each entry $e_{ij}$ corresponds to the logarithm of the relative abundance of the mRNA of gen $g_i$ under a specific condition $c_j$[4]. A bicluster in gene expression data corresponds to the submatrix that genes in that show similar behavior under a subset of conditions. A bicluster is showed by subset of genes and subset of conditions. The similar behavior between genes is measured by *mean squared residue* that was introduced by Cheng and Church.

**Definition 1**: Let $(I,J)$ be a bicluster ($I \subseteq G$, $J \subseteq C$) then the mean squared residue ($r_{IJ}$) of a bicluster $(I,J)$ is calculated as below:

$$r_{IJ} = \frac{1}{|I||J|} \sum_{i\in I, j\in J}(e_{ij} - e_{iJ} - e_{Ij} + e_{IJ})^2 \quad (1)$$

Where

$$e_{iJ} = \frac{\sum_{j\in J} e_{ij}}{|J|} \quad (2)$$

$$e_{Ij} = \frac{\sum_{i\in I} e_{ij}}{|I|} \quad (3)$$

$$e_{IJ} = \frac{\sum_{i\in I, j\in J} e_{ij}}{|I||J|} \quad (4)$$

The lower the mean squared residue, the stronger the coherence exhibited by the bicluster and the quality of the bicluster. If a bicluster has a mean squared residue lower than a given value $\delta$, then we call the bicluster a $\delta$–bicluster. In addition to the mean squared residue, the row variance is used to be relatively large to reject trivial bicluster.

**Definition 2**: Let $(I,J)$ be a biclusters. The row variance of $(I,J)$ is defined as

$$var_{IJ} = \frac{1}{|I||J|} \sum_{i\in I, j\in J}(e_{ij} - e_{iJ})^2 \quad (5)$$

Biclusters characterized by high values of row variance contains genes that present large chances in their expression values under different conditions.

## III. MULTI-OBJECTIVE PARTICLE SWARM

Particle Swarm Optimization (PSO) is a population based on stochastic optimization technique developed by Kennedy and Eberhat in 1995 [5]. This method finds an optimal solution by simulating social behavior of bird flocking. The population of the potential solution is called swarm and each individual solution within the swarm is called a particle. Particles in PSO fly in the search domain guided by their individual experience and the experience of the swarm. Each particle knows its best value so far (pbest) and it's x,y position. This information is an analogy of the personal experience of each particle. More ever each agent knows the best value so far into group (gbest) among pbests. This information is an analog of the knowledge of how the other particles around them have performed.

Each particle tries to modify its position using this information: the current positions $(x_1, x_2, \ldots, x_d)$, the current velocities $(V_1, V_2, \ldots, V_d)$, the distance between the current position and pbest and the distance between the current position and gbest. The velocity is a component in the direction of previous motion (inertia). The movement of the particle towards the optimum solution is governed by updating its position and velocity attributes. The velocity and position update equation are given as [7].

$$V_i^{k+1} = wV_i^k + c_n rand_1 * (gbest_i - s_i^k) + c_2 rand_2 * (gbest - s_i^k) \quad (6)$$

where $v_i^k$ is velocity of agent i at iteration k, w is weighting function, $c_j$ is weighting coefficients, rand is random number between 0 and 1, $s_i^k$ is current position of agent i at iteration k, $pbest_i$ is pbest of agent i, and gbest is gbest of the group.

### A. Binary Particle Swarm Optimization

The binary Particle Swarm Optimization (BinPSO) algorithm was also introduced by Kennedy and Eberhart to allow the PSO algorithm to operate in binary problem spaces [11]. It uses the concept of velocity as a probability that a bit (position) takes on one or zero. In BinPSO, updating a velocity remains the same as the velocity in basic PSO; however, the updating position is redefined by the following rule [11]:

$$s_i^{k+1} = \begin{cases} 0 & if \ r_3 \geq S(v_i^{k+1}) \\ 1 & if \ r_3 < S(v_i^{k+1}) \end{cases} \quad (7)$$

With $r_3 \sim U(0,1)$ and $S()$ is a sigmoid function for transforming the velocity to the probability constrained to the interval [0.0, 1.0] as follows

$$sig(v_i^k) = \frac{1}{1+exp(-v_i^k)} \quad (8)$$

Where $S(v) \in (0,1)$, $S(0)=0.5$, and $r_3$ is a quasi random number selected from a uniform distribution in [0.0, 1.0]. For a velocity of 0, the sigmoid function returns a probability of 0.5, implying that there is a 50% chance for the bit to flip.

### B. particle swarm and Multi-Objective problems

The success of the Particle Swarm Optimization (PSO) algorithm as a single objective optimizer has motivated researchers to extend the use of this bio-inspired technique to Multi-Objective problems [6]. In problems with more than one conflicting objective, there exist no single optimum solution rather there exists a set of solutions which are all optimal involving trade-offs between conflicting objective (pareto optimal set).

**Definition 3**: if there are M objective functions, a solution x is said to *dominate* another solution y if the solution x is no worse than y in all the M objective functions and the solution x is strictly better than y in at least one of the M objective functions. Otherwise the two solutions are *non-dominating* to each other. This concept is shown in Fig1.

**Definition 4**: If Z is subset of feasible solutions, a solution x∈Z is said to *non-dominate* with respect to Z if there does not exist another solution y∈Z that y dominates z (Red point in Fig.1).

**Definition 5**: If F is a set of feasible solutions, a solution x∈F is said to be *pareto-optimal*, if x is *non-dominate* with respect to F (Red point in Fig1 if we suppose all feasible solutions are shown in Fig.1).

In Multi-Objective optimization problem, we determine the pareto optimal set from the set of feasible solutions. In this problem, we must consider the diversity among solutions in pareto set. For maintaining diversity in the pareto optimal set, we use the crowding distance that is provided by deb [12]. In the Multi-Objective PSO, the *nondomiated* solutions are found stored in archive. After each motion in swarm the archive is updated according to bellow:

If an element in archive is dominated by a new solution, the corresponding element in archive is removed. If new solution is not dominated by any element in archive, new solution is added to archive. If archive is full, crowding, distance between elements in archive are computed according to [12] and then one element in archive is selected to remove according to diversity. We use roulette wheel to do this selection.

In (6) each particle need to gbest for motioning in search space. In Multi-Objective PSO we have a set of gbests that called archive. There exists many different ways to select gbest. More detail is described in [6]. In our method, gbest is selected from archives based on crowding distance to maintain diversity. If an element in archive has more diversity, it has more chance to be selected as gbest. We use roulette wheel selection to do it. So the particles motion to pareto optimal set and diversity is maintained with roulette wheel selection for selecting gbest.

## IV. OUR HYBRID MULTI-OBJECTIVE PSO METHOD

Our goal is to find biclusters (I, J) (I is subset of genes, J is subset of conditions) of maximum size, with mean squared residue lower than a given δ, with a relatively high row variance, and with a low level of overlapping among biclusters.

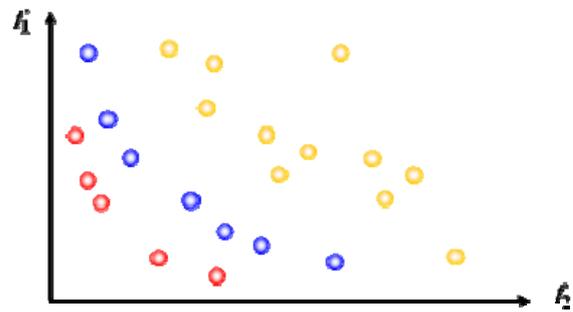

Figure 1. Example of dominate and non-dominate cocepts(f1 and f2 must be minimaze).Red points dominate blue points and yellow points.Red points are non-diminated each other.

The size of bicluster is defined as |I|*|J| if we use this definition as an objective since the number of rows is higher than the number of columns, columns have less effect in objective. So we separate rows and columns and consider two objective functions one for rows and one for columns.

Problem is formulated as below:

Find ( I,J )

That minimize

$$f_1(I,J) = \frac{|G|}{|I|} \quad (9)$$

$$f_2(I,J) = \frac{|C|}{|J|} \quad (10)$$

$$f_3(I,J) = \frac{\delta}{r_{IJ}} \quad (11)$$

$$f_4(I,J) = \frac{1}{var(I,J)} \quad (12)$$

Subject to

$$r_{IJ} \leq \delta \quad (13)$$

In our method, MOPSO with crowding distance is used for solving this problem and it cooperates with local search to find better solution. Since this problem has a constraint (13), we don't apply this constraint when particle move in search space. We allow particle move without any constraint in search space so that they can be stronger to discover new solutions but we add particle to archive if they verity constraint and also a particle is as gbest if constraint is true for it. The problem with overlap among biclusters is addressed in our method as below:

First the archive size equals to 100. After the motion and update of archive, only 50 numbers of particles in the archive that have minimum overlap, move to next generation. So archive with variable size is used. Then in next generation the elements that can be selected as gbest have minimum overlap.

We encode bicluster as particle like [4, 9,13]. Each particle in swarm encodes one bicluster. Biclusters are encoded by means of binary strings of length N+M, where N and M are the number of rows (genes) and number of columns (conditions) respectively. In each particle the first N bits of the binary string are related to genes and the remaining M bits are related to conditions. If a bit is set to 1, it means the related gene or condition belongs to the encoded bicluster; otherwise is does not.

A general scheme of our algorithm is given in figure3. Population is randomly initialized and velocity of each particle in each dimension is set to zero. Then non-dominated population is inserted in archive, after that, we use a local search algorithm to move archive in feasible region. We use Cheng and Church algorithm as local search. The local search algorithm starts with a given bicluster. The irrelevant genes or conditions having mean squared residue above (or below) a certain threshold are now selectively eliminated (or added) using the following conditions [1]. A "node" refers to a gene or a condition. This algorithm contains three phases: multiple node deletion phase, single node deletion phase and multiple node addition phase.

- Multiple nodes deletion :

a) Compute $r_{IJ}, e_{Ij}, e_{iJ}, e_{IJ}$ of the biclusters by (1)–(5).

b) Remove all genes $i \in I$ satisfying

$$\frac{1}{|J|} \sum_{j \in J} (e_{ij} - e_{iJ} - e_{Ij} + e_{IJ})^2 > \alpha * r_{IJ}$$

c) Recompute $r_{IJ}, e_{Ij}, e_{iJ}, e_{IJ}$.

d) Remove all conditions $j \in J$ satisfying

$$\frac{1}{|I|} \sum_{i \in I} (e_{ij} - e_{iJ} - e_{Ij} + e_{IJ})^2 > \alpha * r_{IJ}$$

- Single node deletion,

a) Recompute $r_{IJ}, e_{Ij}, e_{iJ}, e_{IJ}$.

b) Remove the node with largest mean squared residue (done for both gene and condition), one at a time, until the mean squared residue drops below $\delta$.

- Multiple nodes addition.

a) Recompute $r_{IJ}, e_{Ij}, e_{iJ}, e_{IJ}$.

b) Add all conditions $j \notin J$ with

$$\frac{1}{|I|} \sum_{i \in I} (e_{ij} - e_{iJ} - e_{Ij} + e_{IJ})^2 \leq r_{IJ}$$

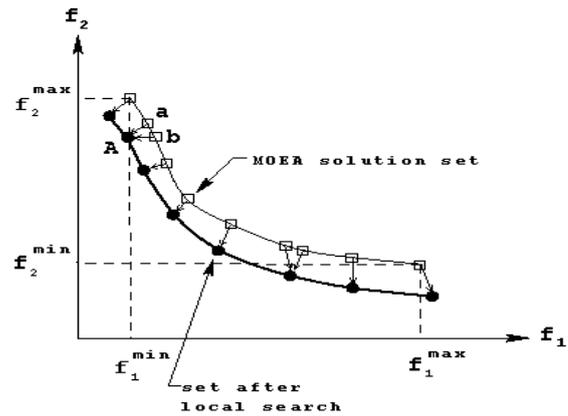

Figure 2. The effects of local search in Multi-Objective optimization(f1 and f2 must be minimaze)

c) Recompute $r_{IJ}, e_{Ij}, e_{iJ}, e_{IJ}$.

d) Add all genes $i \notin I$ with

$$\frac{1}{|J|} \sum_{j \in J} (e_{ij} - e_{iJ} - e_{Ij} + e_{IJ})^2 \leq r_{IJ}$$

This local search is used for particle in the archive. The effects of using local search are illustrated in Fig.2, where the decrement of objective functions is obvious.

```
m=number of column;
n=number of row;
popsize=200;
maxgen=100;
intialize_pop;
intialize_vel;
localsearch_pop();
evaluate;
store_pbests;
insert_nondom;
t=0;
while(t<=maxgen)
  if(nondomCtr > 2)
      compute crowding distance;
  compute_velocity();//according to (7)
  move_particles()//according to (8)
  mutation();
  evaluate();//evaluate objectives value for each particles
  update_archive();
  localSearchArchive();
  update_pbests();
  delete_overlap();// delete overlap particles from archive
  t=t+1;
end while
```

Figure 3. A general scheme of our algorithm

At the beginning, before main loop, these three phases are used but in the main loop after update archive, only the node addition phase is used. In the main loop, crowding is used like [10] to maintain diversity in pareto front.

Three mutation operators are employed: standard mutation operator (a bit is selected randomly and flip), a mutation operator that adds a row and a mutation that adds a column to the bicluster. These three mutations are used with equal probability. In our method the mutation probability for each particle is 0.3.

## V. EXPERIMENTAL RESULT

The proposed biclustering algorithm is implemented in matlab and applied to mine biclusters from two well know data set. The first data set is the *yeast saccharomy cerevisiae cell cycle* expression [1] .The expression matrix contained in this data set consists of 2884 genes and 17 experimental conditions. All entries are integers lying in the range of 0-600. The second data set, the *human B-cells* expression data, is a collection of 4,026 genes and 96 conditions the values of δ for the two data sets are taken from [1]. For the yeast data δ=300 and for the human B-cells expression data δ=1200.

### A. result on yeast data set

Our method is applied to mining fifty biclusters from yeast data set simultaneously this biclusters cover 91.3% of the genes , 100% of the condition and 79.21% cells of the expression matrix while the MOPOB[11] method cover 73.1% of genes and 52.4% cells of the expression data. This improvement might be due to use of variable archive size. In the proposed method, particles move toward the biclusters that have minimum overlaps (gbest). Therefore they may be move towards the cells that have not been covered so far , and the covering facrtor of cells in this algorithm goes higher .In table 1 information about five out of fifty biclusters are summarized.

TABLE I. YEAST BICLUSTERS

| Bicluster | Genes | Conditions | Residue | Row variance |
|---|---|---|---|---|
| 1 | 1079 | 13 | 263.32 | 693.97 |
| 12 | 437 | 17 | 246.85 | 570.66 |
| 37 | 998 | 13 | 275.09 | 730.66 |
| 42 | 808 | 14 | 240.27 | 637.55 |
| 45 | 1660 | 7 | 269.25 | 890.12 |

TABLE II. COMPARATIVE WITH OTHER METHOD FOR YEAST DATASET

| Method | Avg size | Avg residue | Avg genes | Avg condition | Max size |
|---|---|---|---|---|---|
| NAGA 2 | 10301.17 | 234.87 | 109543 | 9.29 | 14828 |
| SEEA 2 B | 8547.21 | 287.56 | 785.42 | 8.92 | 10503 |
| MOPSOB | 10510 | 218.54 | 1102.84 | 9.31 | 15613 |
| OUR Method | 11047 | 259.19 | 12971 | 11.62 | 14027 |

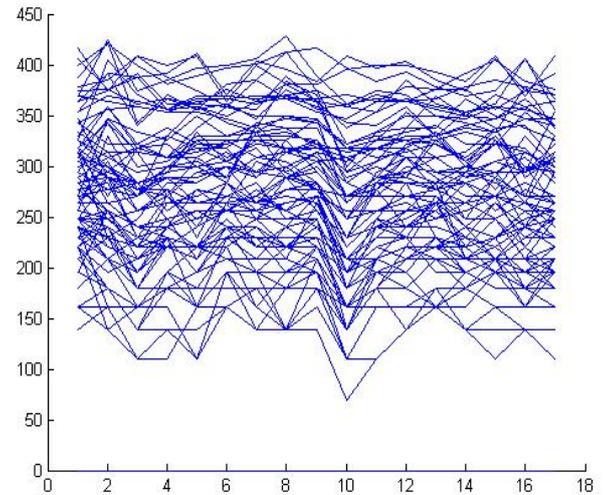

Figure 4. One hunred genes are random selected from bicluster12

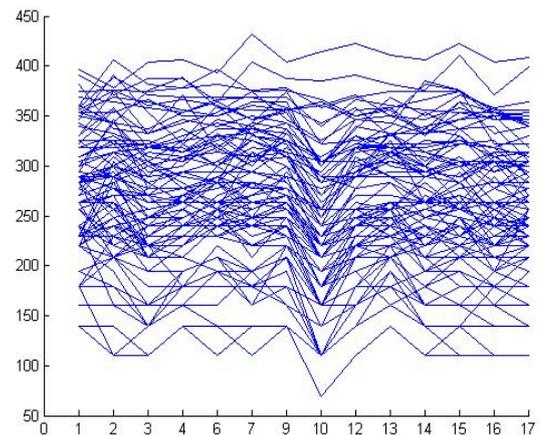

Figure 5. One hunred genes are random selected from bicluster42.

One hundred genes are random selected from bicluster 17and 42 are shown in Fig.4 and Fig.5, respectively.

In order to show the performance of our method, we compare it with other Multi-Objective biclustering method. In [9][13][14] three Multi-Objective biclustering are proposed, we summarize their result, and our result in table 2.

When comparing MOPSOB [11] with our method the average sum square residue is better, but in MOPSOB overlap among bicluster haven't been consider MOPSOB.

### B. result on human data set

Our method is applied to mining one hundred biclusters from human data set too. This biclusters cover 53.6% of the genes , 100% of the condition and 41.6% cells of the expression matrix while the MOPOB[11] method cover 46.7% of genes and 35.7% cells of the expression data. In table 3 information about five out of one hundred biclusters are summarized and a comparative study is expressed in table 4.

TABLE III. HUMAN BICLUSTERS

| Bicluster | Genes | Conditions | Residue | Row variance |
|---|---|---|---|---|
| 1 | 1053 | 36 | 997.68 | 2463.42 |
| 27 | 839 | 42 | 1074.38 | 3570.61 |
| 49 | 105 | 81 | 1197.05 | 2885.34 |
| 73 | 487 | 22 | 769.56 | 5408.31 |
| 92 | 105 | 93 | 1007.41 | 7628.44 |

TABLE IV. COMPARATIVE WITH OTHER METHOD

| Method | Avg size | Avg residue | Avg genes | Avg condition | Max size |
|---|---|---|---|---|---|
| NAGA 2 | 33463.70 | 987.56 | 915.81 | 36.54 | 37560 |
| SEEA 2 B | 29874.8 | 1128.1 | 784.68 | 35.48 | 29654 |
| MOPSOB | 34012.24 | 927.47 | 902.41 | 40.12 | 37666 |
| OUR Method | 33983.64 | 946.78 | 1006.23 | 42.02 | 37908 |

## VI. CONCLUSIONS

In this paper, we introduced an algorithm based on Multi-Objective PSO while incorporating local search for finding biclusters on expression data. In biclustering problem several objective have to be optimized simultaneously. We must find maximum biclusters with lower mean score residue and high row variance. These three objectives are in conflict with each other. We apply hybrid MOPSO and we use crowding distance for maintain diversity. In addition we consider a low level of overlap among biclusters by using archive with variable size. A comparative assessment of results is provided on bench mark gene expression data set to demonstrate the effectiveness of the proposed method. Experimental results show that proposed method is able to find interesting biclusters on expression data and comparative analysis show better performance in result.

Trying other Multi-Objective methods such as the simulated annealing, or employing a neural network in archive clustering can be suggested as future work. Again, decimal encoding of particles may be attempted.

AUTHORS PROFILE

**Seyed Amirhassan Monadjemi**, born 1968, in Isfahan, Iran. He got his PhD in computer engineering, pattern recognition and image processing, from University of Bristol, Bristol, England, in 2004. He is now working as a lecturer at the Department of Computer, University of Isfahan, Iran. His research interests include pattern recognition, image processing, human/machine analogy, and physical detection and elimination of viruses.